\documentclass[a4paper,11pt]{article}
\usepackage{pos}

\def\beq{\begin{equation}}
\def\eeq{\end{equation}}
\def\bea{\begin{eqnarray}}
\def\eea{\end{eqnarray}}

\title{nCTEQ global analysis of nuclear PDFs}

\author*[a]{Michael Klasen}

\affiliation[a]{Institut f\"ur Theoretische Physik, Universit\"at M\"unster, \\ Wilhelm-Klemm-Stra\ss{}e 9, 48149 M\"unster, Germany}

\emailAdd{michael.klasen@uni-muenster.de}

\abstract{We review the series of specific nCTEQ analyses of nuclear parton distribution functions (PDFs) published since 2020 and present preliminary results of a new global analysis. Building on a modern proton baseline without nuclear data and extending the kinematic range, it combines and updates the previous separate analyses that focused on Jefferson Lab neutral-current deep-inelastic scattering (DIS), neutrino DIS and dimuon production, and the currently available CERN LHC data, in particular on W/Z-boson, single inclusive hadron, and heavy-quark production.}

\FullConference{XXXII International Workshop on Deep Inelastic Scattering and Related Subjects (DIS2025)\\
24-28 March, 2025\\
Cape Town, South Africa\\}

\begin{document}
\maketitle

\section{Introduction}

Nuclear parton distribution functions (PDFs) encode important information on the fundamental quark and gluon dynamics of protons and neutrons bound in nuclei. They also determine the initial conditions in the creation of the deconfined quark-gluon plasma in heavy-ion collisions. Their evolution with the resolution scale $Q^2$ can be calculated perturbatively, but the dependence on the longitudinal momentum fraction $x$ must be fitted to experimental data, traditionally from fixed-target deep-inelastic scattering (DIS) or Drell-Yan (DY) lepton-pair production. Over the last years, the field has undergone an enormous development, with pPb collisions at the CERN LHC opening up a wide, previously unexplored regime both in terms of kinematics and processes. 

\section{Previous nCTEQ analyses}

We first briefly review the series of specific nCTEQ analyses published since 2020. Before the advent of CERN LHC data, the nCTEQ15 nuclear PDFs \cite{Kovarik:2015cma}, together with the fits HKN07 \cite{Hirai:2007sx}, EPS09 \cite{Eskola:2009uj}, DSSZ (2012) \cite{deFlorian:2011fp} and nNNPDF1.0 (2019) \cite{AbdulKhalek:2019mzd}, set one of the benchmarks for the global analysis of nuclear PDFs. In three of these analyses (nCTEQ15, EPS09 and DSSZ), fixed-target DIS and DY data were complemented by BNL RHIC pion production data, which provided some constraints on the gluon distribution. The proton PDF baseline in nCTEQ15 was a dedicated CTEQ6.1 fit \cite{Stump:2003yu}, from which the nuclear data from neutrino DIS had been removed. This proton baseline was also used for the series of specific nCTEQ analyses reviewed in this section.

\begin{figure}[b]
 \centering
 \vspace*{-5mm}
 \includegraphics[width=\textwidth]{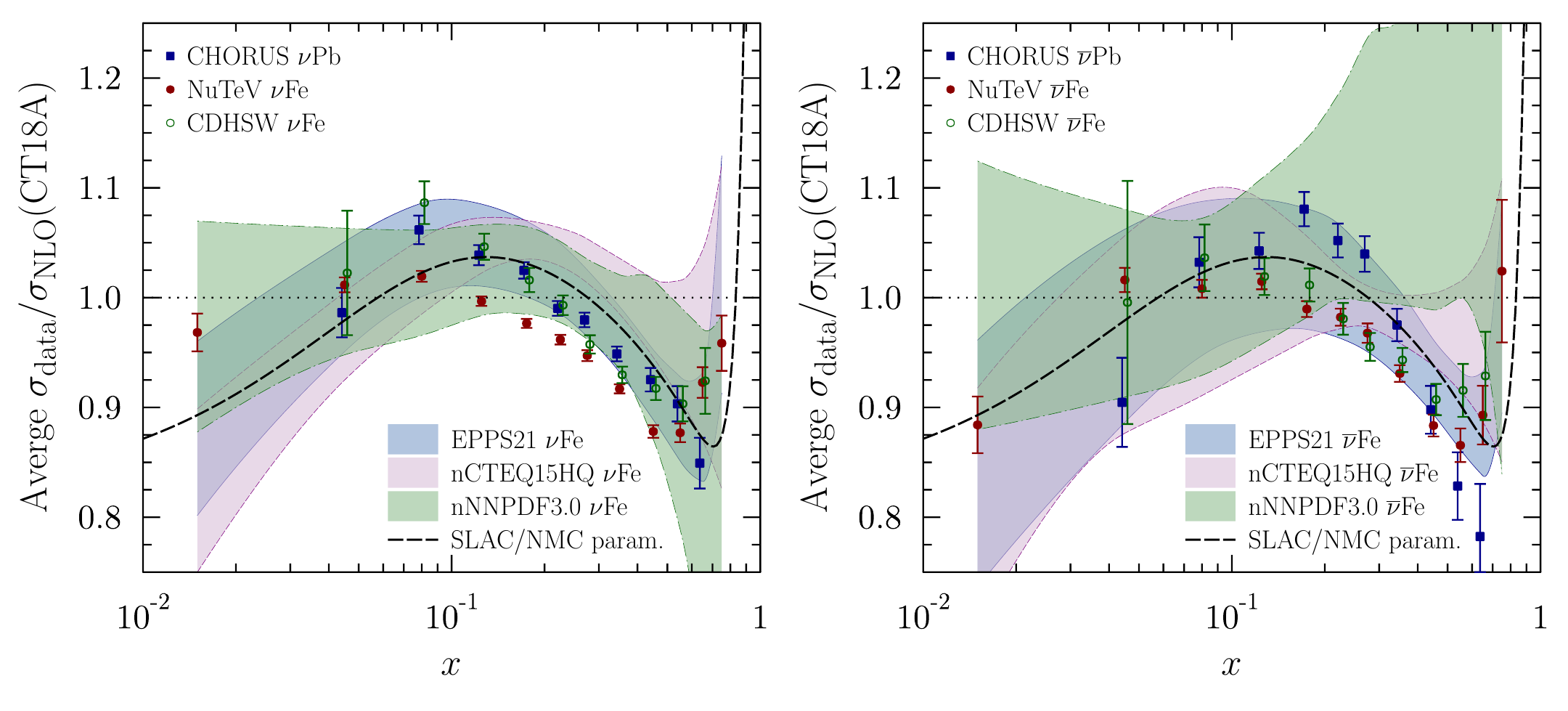}
 \vspace*{-10mm}
 \caption{Average ratios of neutrino (left) and antineutrino (right) cross sections as measured by CHORUS, NuTeV and CDHSW to a theoretical NLO prediction with CT18A PDFs within the kinematic range $Q^2>4$ GeV$^2$ and $W^2>12.25 \,{\rm GeV}^2$. The data are compared with EPPS21 \cite{Eskola:2021nhw}, nCTEQ15HQ \cite{Duwentaster:2022kpv} and nNNPDF3.0 \cite{AbdulKhalek:2022fyi} predictions. The SLAC/NMC parameterisation is also shown as a reference. Taken from Ref.\ \cite{Klasen:2023uqj}.}
 \label{fig:01}
\end{figure}

Historically, the major part of nuclear PDF data has originated from DIS of charged leptons on fixed targets, which is kinematically limited to large $x$ and small $Q^2$. Recent additions to this data set have come from Jefferson Lab. They are significantly more precise than previous data, but lie in a kinematic region which is often excluded by cuts on $Q^2$ and the hadronic mass $W^2$ in global fits. In the nCTEQ15HIX analysis \cite{Segarra:2020gtj}, these cuts were relaxed, which required the inclusion of target mass, higher twist, and deuteron corrections. A good description of the new data could then be obtained with reduced uncertainties at large $x$, in particular for the up and down quark flavours.

\begin{figure}
 \centering
 \includegraphics[width=\textwidth]{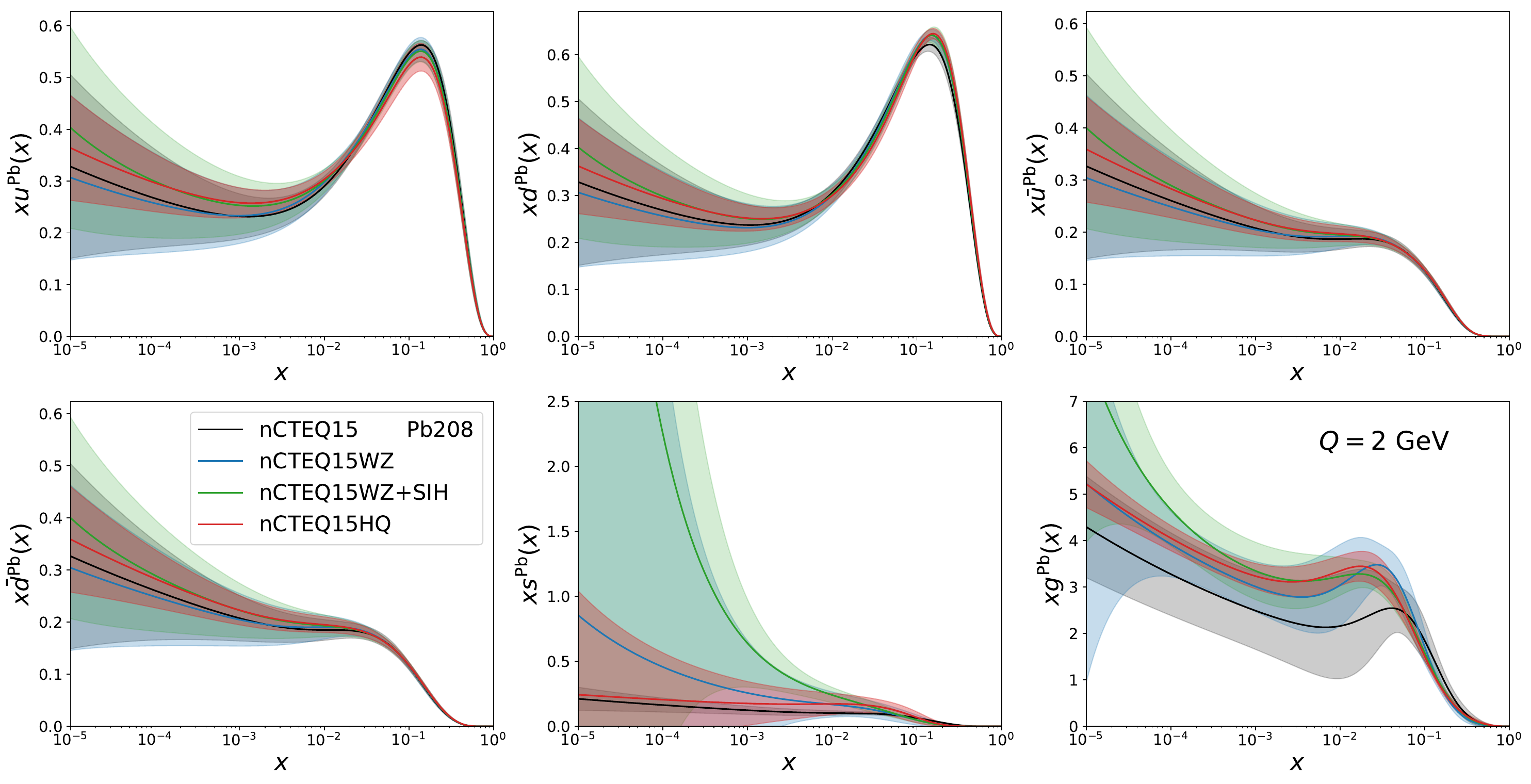}
 \caption{Lead PDFs from recent dedicated nCTEQ analyses. The baseline nCTEQ15 \cite{Kovarik:2015cma} fit is shown in black, nCTEQ15WZ \cite{Kusina:2020lyz} in blue, nCTEQ15WZ+SIH \cite{Duwentaster:2021ioo} in green, and nCTEQ15HQ \cite{Duwentaster:2022kpv} in red.}
 \vspace*{-5mm}
 \label{fig:02}
\end{figure}

A novel, particularly intriguing physical interpretation of this kinematic region of nuclear binding and the EMC effect was proposed in the nCTEQ15SRC analysis \cite{nCTEQ:2023cpo}. Here, $A$-dependent fractions of short-range correlated (SRC) nucleon pairs were fitted together with universal PDFs of nucleons bound in these pairs instead of individual mean-field nucleon PDFs, surprisingly resulting in a lower $\chi^2$. In addition, the fractions of SRC pairs were found to be consistent with results from low-energy nuclear scattering and to be dominated by proton-neutron pairs as expected.

Data on inclusive neutrino DIS have been taken by CDHSW, CCFR and NuTeV on iron and by CHORUS on lead. Also charm quark production has been measured through muonic decays of produced charmed hadrons. These data are only partially included in global analyses due to concerns about possible mutual tensions between the neutrino data sets, tensions with the charged-lepton DIS data, and due to the fact that part of these neutrino data are in some cases already used in the baseline proton PDF fits (cf.\ Fig.\ \ref{fig:01}). Another difficulty is that the data are reported as absolute cross sections and not as ratios to proton or deuteron data. The most consistent results in the nCTEQ15NU analysis were obtained by fitting only charm dimuon and CHORUS data \cite{Muzakka:2022wey}.

The first nCTEQ analysis to include CERN LHC data was nCTEQ15WZ \cite{Kusina:2020lyz}, which took into account all of the available Run-I $W$- and $Z$-boson production data as well as Run-II $W$-boson data from CMS. It was performed with a new code written in C++. Luminosity uncertainties were taken into account by allowing for shifts in the normalisation of the theoretical predictions. In addition to the expected sensitivity to the strange quark, whose contribution was clearly visible, also the gluon distribution was affected through its splitting into sea quarks. In the meantime, ALICE and LHCb have also published new Run-II data on electroweak vector boson production (see below).

\begin{figure}
 \centering
 \includegraphics[width=\textwidth,height=85mm]{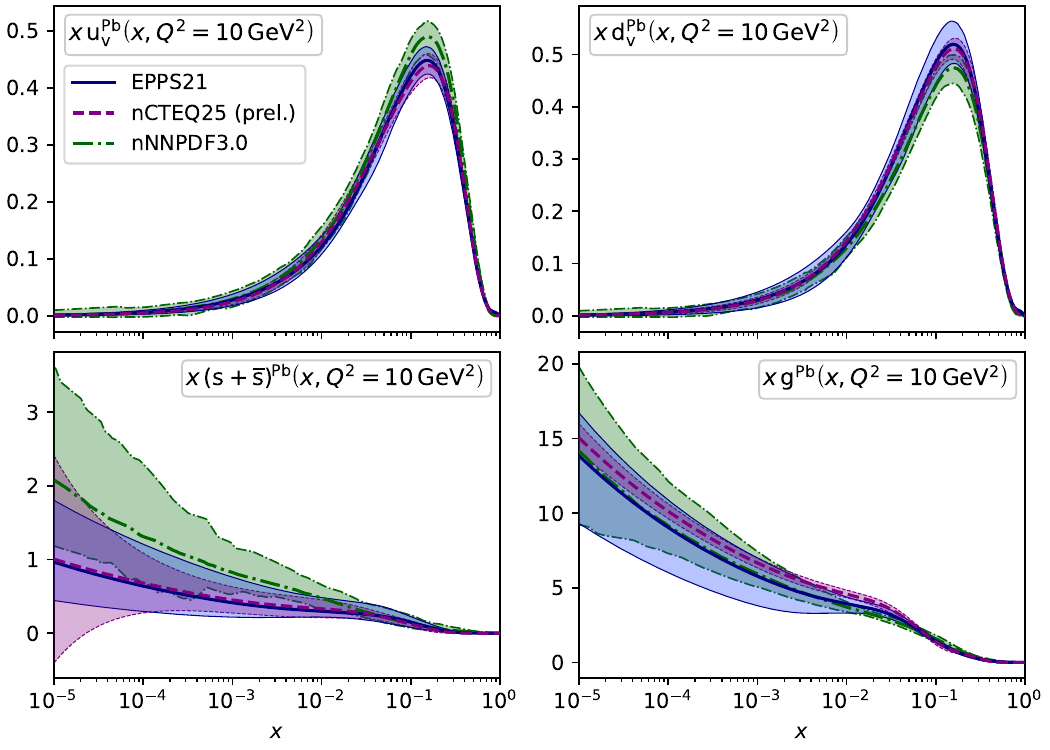}
 \caption{Preliminary nCTEQ25 global analysis results (red, dashed) for valence up (top left), down (top right), strange (bottom left) and gluon (bottom right) nuclear PDFs at $Q^2=10$ GeV$^2$ compared to the corresponding results from the EPPS21 (blue, full) and nNNPDF3.0 (green, dot-dashed) global analyses.}
 \label{fig:03}
\end{figure}

In the nCTEQ15WZ+SIH analysis \cite{Duwentaster:2021ioo}, single-inclusive hadron data from BNL RHIC (PHENIX, STAR) and CERN LHC (ALICE) were added and helped to better constrain the gluon PDF. However, this fit required the inclusion of hadron fragmentation functions, whose impact and uncertainties were therefore thoroughly analysed.

The nCTEQ15HQ analysis \cite{Duwentaster:2022kpv} exploited the vast CERN LHC data set on open heavy-quark and quarkonium production. In this case, hadronisation was taken into account together with the hard matrix element in a data-driven approach with a Crystal Ball function. The latter was extended by an exponential rapidity dependence to account for the forward LHCb kinematics. Due to their abundance and precision, these data significantly improved the precision of the gluon distribution down to previously inaccessibly low values of $x\sim10^{-5}$ (cf. Fig.\ \ref{fig:02}).

\section{Preliminary nCTEQ25 results}

The nCTEQ25 analysis combines the specific analyses described above with the exception of the one using an SRC ansatz. In addition, it takes into account new data that have become available in the meantime, in particular on Run-II W/Z-boson, light and heavy meson production. Theoretical modifications include a new proton baseline (CJ15 \cite{Accardi:2016qay}), a polynomial $x$- and logarithmic $A$-dependence, and a larger number of free parameters and tolerance. Consequently, the fit combines the benefits of the preceding analyses and not only leads to a more precise gluon, but also strange quark distribution (cf.\ Fig.\ref{fig:03}).

\section{Conclusion}

The CTEQ collaboration looks forward to publishing the final nCTEQ25 analysis, which will also include a fully perturbative fit with gridded next-to-leading order (NLO) calculations for open heavy-quark production. Preliminary results show full consistency with the data-driven approach described above.

\section*{Acknowledgment}

The author thanks the organisers for the kind invitation and his nCTEQ colleagues for their collaboration. This work has been supported by the BMBF under contract 05P24PMA.

\end{document}